\documentclass[twocolumn,amsmath,amssymb,nofootinbib]{revtex4}

\usepackage{graphicx}
\usepackage{dcolumn}
\usepackage{bm,psfrag}

\newcommand{\ben}{\begin{equation}}
\newcommand{\een}{\end{equation}}
\newcommand{\be}{\begin{equation}}
\newcommand{\ee}{\end{equation}}
\newcommand{\bea}{\begin{eqnarray}}
\newcommand{\eea}{\end{eqnarray}}
\newcommand{\ba}{\begin{eqnarray}}
\newcommand{\ea}{\end{eqnarray}}

\newcommand{\beq}{\begin{equation}}
\newcommand{\eeq}{\end{equation}}
\newcommand{\beqa}{\begin{eqnarray}}
\newcommand{\eeqa}{\end{eqnarray}}
\newcommand{\beqar}{\begin{eqnarray*}}
\newcommand{\eeqar}{\end{eqnarray*}}

\newcommand{\reef}[1]{(\ref{#1})}

\newcommand{\eg}{{\it e.g.,}\ }
\newcommand{\ie}{{\it i.e.,}\ }
\newcommand{\comment}[1]{{\bf [[[#1]]]}}

\newcommand{\labell}[1]{\label{#1}} 

\newcommand{\cO}{{\cal O}}

\def\t6 {T_\mt{D6}}

\newcommand{\mt}[1]{\textrm{\tiny #1}}

\newcommand{\vk}{{\vec{k}}}

\def\cale         {{\cal E}}

\def\calo         {{\cal O}}

\def\del          {\partial}

\def\ee           {{\rm e}}

\def\sqr#1#2{{\vcenter{\vbox{\hrule height.#2pt
 \hbox{\vrule width.#2pt height#1pt \kern#1pt
 \vrule width.#2pt}\hrule height.#2pt}}}}

\def\w{\omega}

\def\ee{\cale}

\def\aa1{\phi}
\def\cc1{\psi}

\def\comment#1{{\bf [[#1]]}}

\newcommand{\dg}{\delta \lambda}
\newcommand{\dt}{\delta t}
\newcommand{\dE}{\delta{\cal E}}

\usepackage[bookmarks=false]{hyperref} 
\hypersetup{pdfstartview=FitH,pdfhighlight=/O,colorlinks=false}

\begin{document}

\preprint{arXiv:1312.nnnn [hep-th]; UWO-TH-13/xx}

\title{Universal scaling in fast quantum quenches in conformal field theories}

\author{Sumit R. Das,$^{1}$ Dami\'an A. Galante$^{2,3}$ and Robert C. Myers$^3$}
\affiliation{$^1$\,Department of Physics and Astronomy, University of Kentucky, Lexington, KY 40506, USA}
\affiliation{$^2$\,Department of Applied Mathematics, University of Western Ontario, London, Ontario N6A 5B7, Canada}
\affiliation{$^3$\,Perimeter Institute for Theoretical Physics, Waterloo, Ontario N2L 2Y5, Canada}


\begin{abstract}
We study the time evolution of a conformal field theory deformed by a
relevant operator under a smooth but fast quantum quench which brings
it to the conformal point. We argue that when the quench time scale
$\delta t$ is small compared to the scale set by the relevant
coupling, the expectation value of the quenched operator scales
universally as $\dg/ \dt ^{2\Delta-d}$ where $\dg$ is the quench
amplitude. This growth is further enhanced by a logarithmic factor in
even dimensions. We present explicit results for free scalar and
fermionic field theories, supported by an analytic understanding of
the leading contribution for fast quenches. Results from this Letter
suggest that this scaling result, first found in holography, is in
fact universal to quantum quenches. Our considerations also show that
this limit of {\em fast smooth} quenches is quite different from an
{\em instantaneous} quench from one time-independent 
Hamiltonian to another,
where the Schrodinger picture state at the time of the quench simply
serves as an initial condition for subsequent evolution with the final
Hamiltonian.
\end{abstract}

\maketitle

\noindent \textbf{1. Introduction:}
Quantum quenches involve changing one of the parameters of
a quantum system and then following its subsequent evolution.
Recently, a great deal of attention has been given to the study of such systems, mainly because they have become available in laboratory experiments \cite{more}.  A particularly interesting class of quenches are those which involve a critical point
at either an initial, final or intermediate stage of the quench. In this case, one expects that the subsequent time evolution 
carries universal signatures of the critical point.
Indeed, for quenches which start in an initially gapped phase, it has been conjectured that for {\em slow} quenches various physical quantities obey scaling properties in the critical region known as Kibble-Zurek scaling \cite{kibblezurek,more} which involves the equilibrium and dynamical critical exponents. At the other extreme,
for instantaneous quenches to a critical point, Calabrese and Cardy \cite{cc2,cc3} demonstrated various universal characteristics and obtained exact results in two dimensions using methods of boundary conformal field theory. Further scaling relations for the fidelity susceptibility were found by studying instantaneous quenches with
small amplitude \cite{gritsev}.

The study of quenches can also produce new insights for questions related to understanding the development of thermal
behaviour. Motivated largely by the study of thermalization of the strongly coupled quark-gluon plasma \cite{holo-therm}, 
quenches have become an active topic of study within the framework of gauge/gravity duality.
The AdS/CFT correspondence provides a particularly useful tool to analyze these kind of processes, since it allows the possibility of studying strongly coupled theories, to follow the real time evolution of the system and to easily vary other parameters such as the space-time dimension and/or the temperature.
Indeed, in recent years holographic methods have thrown valuable insight into issues of thermalization \cite{holo-therm}, universal Kibble-Zurek behavior in critical dynamics \cite{holo-slow} and the dynamics of relaxation following quench across a critical point \cite{holo-bhaseen}.

In \cite{numer,fastQ}, a new set of scaling properties in the early time behaviour
were found in the holographic analysis of {\em smooth} but {\em fast} quantum quenches where a critical theory was deformed by a relevant operator ${\cal O}_\Delta$, with conformal dimension $\Delta$, with a time dependent coupling $\lambda(t)$. Specifically, if $\dg$ denotes the amplitude of the quench and $\dt$ is the time scale of the quench duration, it was found that for fast quenches (\ie when $\dg (\delta t)^{d-\Delta} \ll 1$), the change in the (renormalized) energy density $\dE$ scales as $\dg^2 /\dt ^{2\Delta-d}$ and the peak in the (renormalized) expectation value of the operator $\langle\cO_\Delta\rangle$ scales as $\dg /\dt ^{2\Delta-d}$. In fact, the growth in the expectation value is enhanced by
an additional logarithmic factor for even $d$ if the $\Delta$ is an integer and for odd $d$ if $\Delta$ is a half integer.
The same scaling holds for a reverse quench, \ie a quench from a gapped theory to the critical theory. Note that in the limit $\dt\rightarrow 0$, these scalings yield a physical divergence when the conformal dimension is greater than $d/2$.

The divergences noted above for the limit $\dt\to0$ seem to imply that infinitely fast quenches cannot be physically realized.
This conclusion is clearly at odds with the extensive studies of these processes, \eg \cite{cc2,cc3,gritsev}, where an `{\em instantaneous} quench'
is introduced as the starting point of the analysis. In these studies, there is an instantaneous transition from an initial (time-independent) hamiltonian $H_i$ to a final (time-independent) hamiltonian $H_f$ at a time $t_0$. Then an initial state set
at $t\to-\infty$ evolves to some state $|\psi_0\rangle$ at $t=t_0$ by evolution with hamiltonian $H_i$. This latter state then acts an initial state in the time evolution for $t > t_0$ with the new hamiltonian $H_f$. In contrast, the smooth fast quenches we consider involves time evolution with a time dependent hamiltonian which is a smooth function of time. A significant aspect of our work is to understand how the fast limit of a smooth quench generally produces quite different behaviour from the
instantaneous quenches considered in \cite{cc2,cc3,gritsev}. 

Of course, there is no a priori reason to expect that the results in these two analyses should be similar. The holographic studies deal with a rather special class of  strongly-coupled large-$N$ CFTs, that are clearly different from, \eg the weakly coupled field
theories considered in \cite{cc2,cc3}.  However, in this Letter we will argue that the scaling results of \cite{numer,fastQ} for fast but smooth quenches are in fact quite general. We first support this claim by demonstrating that the same scaling behaviour arises even in quenches of simple free field theories. Further we are able to argue that the same scaling result holds for a broad class of quenches of CFT's.

\noindent \textbf{2. Quenching a free scalar field:}  We begin here by analyzing mass quenches for a free scalar field $\phi$
in a general spacetime dimension $d$. In particular, we consider a time-dependent mass which makes a smooth transition from some nonzero value $m$ at early times to zero at asymptotically late times, with a profile which allows an exact solution for arbitrary quench rates. In the limit, $\dt\to0$, the profile yields an instantaneous transition from $m$ to zero and so
allows for a direct comparison to the results of \cite{cc2,cc3}. However, our analysis is guided by the holographic analysis of \cite{numer,fastQ}. From this perspective, the time-dependent coupling is $\lambda(t)=m^2(t)$ while the operator is
$\cO_\Delta=\phi^2$ with conformal dimension $\Delta=d-2$.  In the fast quench limit, we will demonstrate that the same scaling behavior discovered in the holographic analysis \cite{numer,fastQ} appears here for the suitably renormalized expectation value, $\langle\cO_\Delta\rangle_{ren}\sim\dg/\dt ^{2\Delta-d}=
\dg/\dt ^{d-4}$. In fact, our analytic results also reveal an additional logarithmic enhancement of this scaling for
even $d$, which is again in agreement with the holographic analysis. Further, comparing our analysis to \cite{cc2,cc3}, 
we show that our momentum-space correlation functions reproduce the results of an instantaneous quench in \cite{cc3} only when the time scale for quench is small compared to all of the relevant momenta, as well as being small compared to the initial mass.

Our choice for the profile of the mass is 
\begin{eqnarray}
m^2(t) = \frac{m^2}{2} \left(1 - \tanh {t}/{\dt}\, \right)\,,
\label{massfunction}
\end{eqnarray}
that interpolates between $m^2(t=-\infty)= m^2$ and $m^2(t=+\infty) = 0$. Hence we start in a massive theory and we end with the massless case. Luckily, the
scalar field equation,  
\beq
\left(\Box + m^2(t)\right)\,\phi=0\,,
\label{eom}
\eeq
with this profile \reef{massfunction}, was studied previously as an example of quantum fields in a cosmological background \cite{BD,BD2}. Hence exact analytical solutions for the mode functions are known with
\begin{eqnarray}
\phi= \int\!\! d^{d-1}k\ \left( a_\vk\, u_\vk + a^\dagger_\vk\, u^*_\vk\right)\,,
\labell{fieldx}
\end{eqnarray}
where $u_\vk$ are the in-modes
\begin{eqnarray}
&&\!\!\!\!
u_k  = \frac{1}{\sqrt{4\pi \omega_{in}}} \exp(i\vk\cdot\vec{x}-i\omega_+ t - i\omega_- \dt \log (2 \cosh t/\dt)) \nonumber\\ 
&&\!\!\!\!
\times\, _2F_1 \left( 1+ i \omega_- \dt, i \omega_- \dt; 1 - i \omega_{in} \dt; \frac{1+\tanh(t/\dt)}{2} \right)
\label{modes}
\end{eqnarray}
with $\omega_{in}  =  \sqrt{\vk^2+m^2}$ and
$\omega_{\pm}  = (|\vk|\pm\omega_{in})/2.$
The operators $a_\vk$ above are defined to annihilate the in-vacuum, $a_\vk |in,0\rangle=0$.
Now it is straightforward to compute 
the expectation value 
\begin{eqnarray}
\langle\phi^2\rangle\equiv
\langle in,0|\phi^2|in,0\rangle = \frac{1}{2(2\pi)^{d-1}}\int \frac{d^{d-1}k}{\omega_{in}} |_2F_1|^2\,.
\label{phi_squared}
\end{eqnarray}

Of course, this expectation value \reef{phi_squared} contains UV divergences associated with the integration of
$k=|\vk|\to\infty$. The standard approach to deal with these UV divergences is to add suitable counterterms involving
the time-dependent mass to the effective action, as in the holographic renormalization of \cite{numer}. 
The approach is in fact identical to renormalization of quantum field theories in curved space-time, with the time-dependent mass playing the role of a background metric. In this latter case, the counterterms consist of all diffeomorphism invariant operators upto dimension $d$ which can be constructed from the field and its derivatives, as well as geometric tensors constructed from the background metric. In the present case, this means that we should add counterterms which involve powers of the mass as well as time derivatives of the mass.
While we will
describe this approach in more detail for the present quenches in \cite{future}, here it is sufficient to identify the
contributions of these counterterms which render eq.~\reef{phi_squared} finite. Hence let us write the renormalized
expectation value as
\begin{eqnarray}
\langle\phi^2\rangle_{ren} \equiv \frac{\Omega_{d-2}}{2(2\pi)^{d-1}} \int dk \left( \frac{k^{d-2}}{\w_{in}} |_2F_1|^2 - f_{ct}(k,m(t)) \right)\,. \label{renorm}
\end{eqnarray}
where $f_{ct}(k,m(t))$ designates the counterterm contribution and $\Omega_{d-2}$ denotes the angular volume of a unit
($d$--2)-dimensional sphere.

The counterterm contributions are readily identified by considering slow quenches and employing an adiabatic expansion \cite{BD2}, as we now sketch. First, we define the `adiabatic' modes as
\begin{eqnarray}
u_\vk = \frac{1}{\sqrt{4\pi \Omega_k}} \exp \left(i \, \vk\cdot\vec{x} -i \int^t \Omega_k(t') dt' \right)
\end{eqnarray}
for which the (bare) expectation value \reef{phi_squared} becomes
 \beq
\langle\phi^2\rangle =  \frac{\Omega_{d-2}}{2(2\pi)^{d-1}} \int dk  \frac{k^{d-2}}{\Omega_k}\,.
\label{bare1}
 \eeq
In order for these modes to satisfy the Klein-Gordon equation \reef{eom} with a time varying mass, $\Omega_k$ 
satisfies
\begin{eqnarray}
\Omega_k^2 = k^2 + m^2(t) - \frac{1}{2}  \frac{\partial_t^2{\Omega}_k}{\ \Omega_k} + \frac{3}{4} \left(\frac{\partial_t{\Omega}_k}{\Omega_k} \right)^2 \,. \label{Omega}
\end{eqnarray}
Now the adiabatic analysis is applied by solving eq.~\reef{Omega} with the assumption that time-derivatives of the mass are small, \ie $\partial_t^n m\ll m^{n+1}$. Given this solution, one can identify the UV divergences in eq.~\reef{bare1} by expanding
the integrand for large $k$, \ie $\partial_t^n m\ll k^{n+1}$. The final result takes the form
\begin{eqnarray}
&&\!\!\!\!
f_{ct}(k,m(t)) =  k^{d-3} - \frac{k^{d-5}}{2} m^2(t) + 
\frac{k^{d-7}}{8} \left( 3m^4(t) \right.
\nonumber\\
&& \qquad\qquad +\left.\partial^2_t m^2(t)\right) 
 - \frac{k^{d-9}}{32} \left(10 m^6(t) + \partial^4_t m^2(t)
\right.\label{ict}\\
&&\qquad\qquad \left.+ 10 m(t)^2\, \partial^2_t m(t)^2+5\partial_t m(t)^2\, \partial_t m(t)^2\right)  + \cdots \,.
\nonumber 
\end{eqnarray}
The terms above should only be included only if the power of $k$ greater or equal to $-1$. Hence eq.~\reef{ict} shows
the counterterm contributions needed to renormalize the expectation value \reef{renorm} up to $d=9$. 
Let us emphasize that here we are
using the adiabatic limit simply as a convenient approach to identify the counterterm contributions, however, these same terms \reef{ict} still renormalize the expectation value \reef{renorm} for arbitrary mass profiles and, in particular, for quenches with $\dt\to0$.
The physical intuition behind this expectation is that the very high momentum modes should be insensitive to the quench rate, \ie
changing $\dt$ does not affect physics at the regulator scale $\Lambda$, as long as $\Lambda\dt \gg 1$.

As an independent check, we have verified that precisely the same counterterm contributions \reef{ict} appear in
directly making a large $k$ expansion in eq.~\reef{phi_squared}. First we define dimensionless parameters,
$\kappa = m\delta t$ and $q = k \delta t$, and then expand for small $\kappa$ at fixed $q$ in 
the series representation of the hypergeometric functions
\begin{eqnarray}
_2F_1 (a,b;c;z)  = \sum_{n=0}^{\infty} \frac{(a)_n (b)_n}{(c)_n} \frac{z^n}{n!},
\end{eqnarray}
where $(x)_n = x (x+1) \cdots (x+n-1)$ (and $(x)_0 = 1$) and with $a,b,c,z$ defined as in eq.~(\ref{modes}). 
Now while every term in the above series has a $\kappa^2$ contribution, only the first few terms contribute in eq.~\reef{phi_squared} after integrating. The UV divergent terms in the integral are removed as in eq.~\reef{renorm} by
subtracting a few extra terms with definite powers of $q$. The coefficients of these terms have a complicated
time dependence, however,  we have verified that they precisely match the expressions in eq.~\reef{ict} for the 
given mass profile \reef{massfunction}.

Though the above discussion applies for a general space-time dimension, we should distinguish between odd and even dimensions. When $d$ is even, apart from the usual power-law divergences, we also find logarithmic divergences, \ie
eq.~\reef{ict} contains a $k^{-1}$ term. Hence, for even $d$, we also need to add a renormalization scale $k_0$ in
defining the renormalized expectation value \reef{renorm}.  The appearance of this scale reflects new scheme-dependent ambiguities which can arise with time-dependent couplings \cite{numer}. Taking account of this additional complication,
we are now able to compute the expectation value of $\langle\phi^2\rangle_{ren}$ for any dimension.

\begin{figure}[h]
\begin{center}
 \includegraphics[scale=0.65]{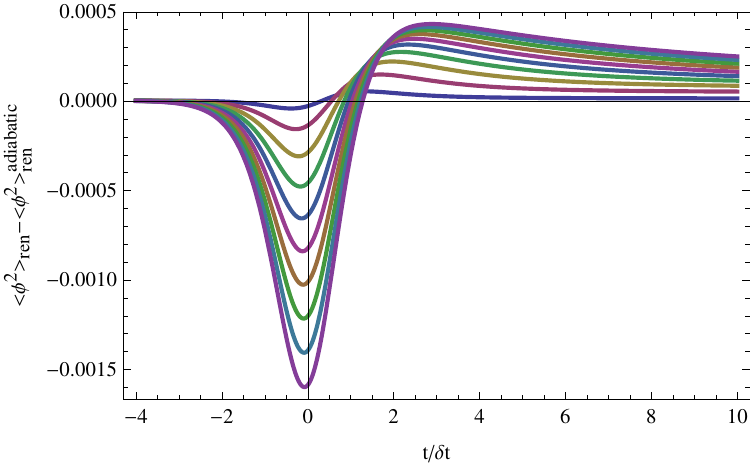}
\end{center}
  \caption{(Colour online)  Expectation value $\langle\phi^2\rangle_{ren}$ as a function of $t/\dt$
for different values of $\dt$ with $d=5$. The
values of $\dt$ correspond to $1, 1/2, 1/3, \cdots, 1/10$, with their peaks increasing as $\dt$
becomes smaller.} \label{fig_d5}
\end{figure}

\begin{figure}[h]
\begin{center}
 \includegraphics[scale=0.45]{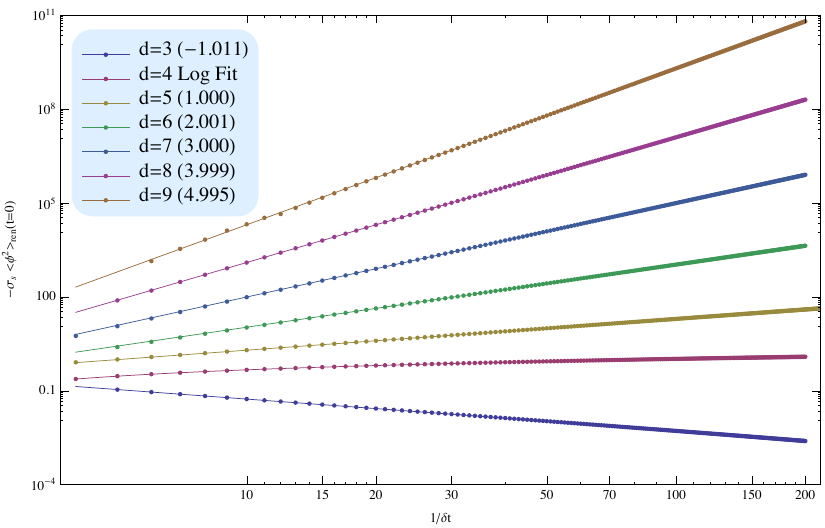}
\end{center}
  \caption{(Colour online) Expectation value $\langle\phi^2\rangle_{ren}(t=0)$ as a function of the quench times $\dt$ for space-time dimensions from $d=3$ to $d=9$. Note that in the plot, the expectation values are multiplied by a
numerical factor depending on the dimension: $\sigma_s = 2 (2\pi)^d/\Omega_{d-2} $.
The slope of the linear fit in each case is shown in the brackets beside the labels.
The results  support the power law relation $\langle\phi^2\rangle_{ren} \sim \dt^{-(d-4)}$.} \label{alld}
\end{figure}

It is straightforward to evaluate eq.~\reef{renorm} numerically. As an example, figure \ref{fig_d5}
illustrates the result for $d=5$ for various values of $\dt$. Certainly, we see in the figure that the peak of the expectation
value grows as $\dt$ becomes smaller. To evaluate this growth quantitatively, consider a log-log
plot of the expectation value at $t=0$ versus $\dt$, \ie $\langle\phi^2\rangle_{ren}(t=0)$ as a function of
$\dt$, in figure \ref{alld}. The slope of a linear fit to this data is very close to one, indicating $\langle\phi^2\rangle_{ren}(t=0)
\propto 1/\dt$ for $d=5$. In fact, we considered spacetime dimensions $9\ge d\ge 3$ in figure \ref{alld} and our
numerical results indicate that generally $\langle\phi^2\rangle_{ren}$ scales as $\dt^{-(d-4)}$,  which matches
the scaling behaviour uncovered by the holography analysis \cite{numer,fastQ}. Note that for $d=4$, the naive formula suggests
that  $\langle\phi^2\rangle_{ren}(t=0)$ is independent of $\dt$,  but instead we found a logarithmic scaling, which again matches the holographic results. The holographic
analysis \cite{numer,fastQ} further suggests that the power-law growth should be enhanced by an extra logarithmic factor
for general even $d$. Unfortunately, we found that the fits to our numerical results are only sensitive to this logarithmic enhancement for $d=4$, in which case, it actually provides the leading scaling behaviour --- however, see our analytic results below. Finally, it
is noteworthy that the above scaling continues to be valid for $d=3$, where we find that the expectation value scales as $\dt$, 
\ie $\langle\phi^2\rangle_{ren}(t=0)$ vanishes rather than diverging as $\dt \rightarrow 0$. This case is examined more carefully in \cite{future}.

\begin{figure}[h]
\begin{center}
 \includegraphics[scale=0.8]{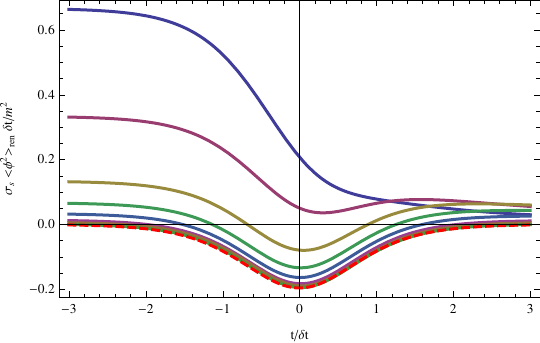}
\end{center}
  \caption{(Colour online) Scaled expectation value $\dt/m^2\times \langle\phi^2\rangle_{ren}$ for $d=5$. 
At early times, from top
to bottom, the solid curves correspond to $\dt=1$, 1/2, 1/5, 1/10, 1/20, 1/50, 1/100, 1/500. As $\dt$ gets smaller, the curves approach the leading order solution \reef{scaling1} (shown with the dashed red line) found with the $\kappa$ expansion. The last solid curve corresponding to $\dt=1/500$ essentially
matches this analytic result.} \label{d5_an}
\end{figure}
In addition, we note that, in fact, it is not only the peak value of $\langle\phi^2\rangle_{ren}$ that
exhibits the scaling observed above. Rather in the limit of fast quenches, the entire response approaches a universal form 
with this scaling, as illustrated in figure \ref{d5_an}. This figure shows the expectation value scaled by
$\dt^{d-4}/m^2$ as a function to $t/\dt$ for the case $d=5$. We see that the full response collapses down
towards a single universal curve as $\dt\to 0$. In fact, extending the $\kappa$ expansion described above, we 
can obtain an analytic expression for this universal behaviour in
the $\dt\to0$ limit. For odd $d\geq5$, we find the leading contribution takes the form
\begin{eqnarray}
\langle\phi^2\rangle_{ren} = \frac{\Omega_{d-2}}{2(2\pi)^{d-1}} (-1)^{\frac{d-1}{2}} \frac{\pi}{2^{d-2}}\, \partial^{d-4}_t m^2(t) + O(\dt^{-(d-6)}). \nonumber \\
\label{scaling1}
\end{eqnarray}
As illustrated in fig.~\ref{d5_an}, this expression for the leading order contribution
reproduces the time-dependence of $\langle\phi^2\rangle_{ren}$ found numerically for 
very small $\dt$.  The general scaling discovered previously can be seen by substituting in the
given mass profile (\ref{massfunction}), in which case eq.~\reef{scaling1} yields
\ben
\langle\phi^2\rangle_{ren} \sim m^2/ (\delta t)^{d-4}
\label{scaling}
\een
as the general scaling at leading order in the $\kappa$ expansion. Of course, this result matches the scaling of $\langle\phi^2\rangle_{ren}(t=0)$ found numerically for small $\dt$ and 
it is also the scaling behavior obtained in the holographic calculations for a strongly coupled theory \cite{numer,fastQ}. 
The case of $d=3$ can also be treated separately and we obtain
\bea
\langle\phi^2\rangle_{ren} = \frac{\Omega_{d-2}}{2(2\pi)^{d-1}} \frac{\pi}{4} m^2 \delta t \log \left(\frac{1-\tanh{t/\delta t}}{2}\right)\,,
\eea 
to leading order in $\dt$.

For even $d$, the situation is similar but we find analytically the dependence with the renormalization
scale $k_0$. The final result is
\begin{eqnarray}
\langle\phi^2\rangle_{ren} = \frac{\Omega_{d-2}}{2(2\pi)^{d-1}} \frac{(-1)^{d/2}}{2^{d-3}}\, \log(1/k_0\dt)\, \partial^{d-4}_t m^2(t) + \cdots, \nonumber \\
\end{eqnarray}
where the `dots' indicate terms independent of $k_0$. Again, this formula reproduces the scaling behaviour
originally found with holography, including an extra logarithmic enhancement, \ie $\langle\phi^2\rangle_{ren} \sim 
\log(1/\dt)/\dt^{d-4}$.

As an independent check of our results, we have considered the diffeomorphism Ward identity \cite{numer} 
\begin{eqnarray}
\partial_t \langle {\cal E}\rangle=-\langle {\cal{O}}_\Delta\rangle\ \del_t \lambda\,,
\labell{ward}
\end{eqnarray}
where $\cal E$ is the energy density. The energy density may be evaluated independently \cite{future} and comparing
the results for the two sides of the above identity shows complete agreement. We also note that these calculations
verify the quenches of the free scalar reproduce the scaling  $\langle {\cal E}\rangle\sim\dg^2 /\dt ^{2\Delta-d}
\sim(m^2)^2 /\dt ^{d-4}$,
which was found in the holographic analysis \cite{numer,fastQ}.

Hence our analysis indicates that both $\langle\phi^2\rangle_{ren}$ and $\langle {\cal E}\rangle$
diverge in the limit $\dt\rightarrow 0$ for $d\ge 4$.
These divergences suggest that infinitely fast quenches 
are physically inconsistent, which creates a clear tension with
previous studies where an `{\em instantaneous quench}'
forms the starting point of the analysis. Hence we close this section by 
reconciling our results above with those for the
instantaneous quenches studied in \cite{cc2,cc3}. 
To begin, consider the momentum-space correlation function $\langle\phi(\vk,t)\,\phi(-\vk,t^\prime)\rangle$. This is equivalent to 
the correlation function of a single harmonic oscillator with a time-dependent frequency $\omega^2(t) = k^2 + m^2(t)$ and the corresponding correlator appears in eq.~(8) of \cite{cc3}
for an {\em instantaneous} quench.
In the present framework with finite $\dt$, the corresponding correlator is easily calculated using the Bogoliubov coefficients between the ``in" and ``out" modes for this problem \cite{BD,BD2}. 
In fact, we find that the resulting correlation function for the smooth quenches 
reproduces precisely that given in \cite{cc3} for a
particular $\dt\to0$ limit, where $k \delta t \ll 1$,  $m\delta t \ll 1$ and $\dt\ll
t, t^\prime$ \cite{future}. Hence in this respect, the infinitely fast limit of our smooth quenches agrees with
the instantaneous quenches of \cite{cc2,cc3}.

However, a departure between the two approaches becomes apparent when considering
correlation functions in real-space. To obtain any real-space correlator, 
we need to integrate the momentum-space correlators over all momenta (up to some cutoff $\Lambda$).
Hence in our approach with finite $\dt$, this integration will include contributions of momentum-space
correlators with $k \dt \gtrsim 1$ whereas agreement with the instantaneous quench for these
correlators demands that $k \delta t \ll 1$. Therefore, the instantaneous quench results for real-space correlators  \cite{cc2,cc3} will not appear as the limit of our results. However, one may still expect that both approaches will yield the same results for certain infrared questions, \eg late time correlators for separations large compared to $\delta t$ \cite{future}. 
We should 
add that much of \cite{cc2,cc3} focuses on lower dimensions, \eg
$d=2,3$, where our analysis does not reveal any divergences for the free scalar in the $\dt\to0$ limit.

\noindent \textbf{3. Quenching a free fermionic field:} Here we turn to mass quenches for a free Dirac fermion $\psi$
in a general spacetime dimension $d$. The calculations fit these quenches are completely analogous to those above for
the scalar field and so we only provide a sketch here, reserving the details for \cite{future}. Applying the terminology of the
holographic studies \cite{numer,fastQ} here, the time-dependent coupling is $\lambda(t)=m(t)$ while the operator is
$\cO_\Delta=\bar\psi\psi$ with conformal dimension $\Delta=d-1$.  Hence the scaling behavior expected from the holographic results \cite{numer,fastQ} would be $\langle\cO_\Delta\rangle_{ren}\sim\dg/\dt ^{2\Delta-d}=\dg/\dt ^{d-2}$. 

\begin{figure}[h]
\begin{center}
 \includegraphics[scale=0.55]{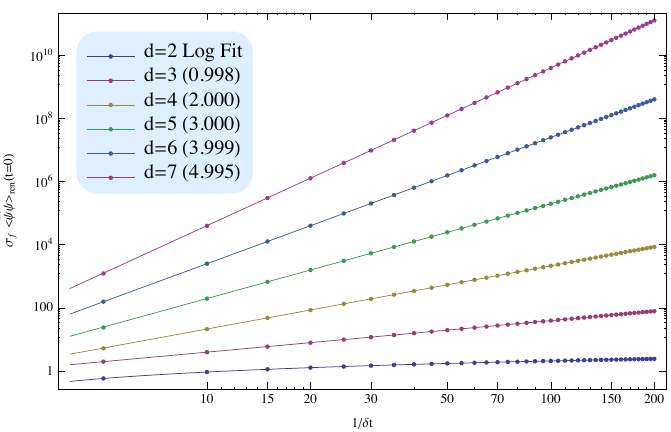}
\end{center}
  \caption{(Colour online) Expectation value $\langle\bar{\psi}\psi\rangle_{ren}(t=0)$ as a
function of the quench times $\dt$ for space-time dimensions from $d=2$ to $d=7$. Note that in the plot, the expectation values are multiplied by a
numerical factor depending on the dimension: $\sigma_f = (2\pi)^{(d-1)/2}/\Omega_{d-2}$. The slope of the linear fit in each case is shown in the brackets beside the labels.
The results support the power law relation $\langle\bar{\psi}\psi\rangle_{ren} \sim \dt^{-(d-2)}$.} \label{fig_alld_f}
\end{figure}

As for the scalar, the present fermionic quenches can be related to fermions (with constant mass) in a time-dependent cosmological background \cite{Duncan}. In particular, with the mass profile
\begin{eqnarray}
m(t) = \frac{m}2\left(1 - \tanh t/\dt\,\right)\,,
\end{eqnarray} 
analytic solutions for the mode functions can be found.  The expectation value $\langle\bar\psi\psi\rangle$ must again
be renormalized by subtracting various counterterm contributions, as above.  The resulting expression is easily
evaluated numerically and the results are shown in figure \ref{fig_alld_f} for various $d$. In particular, the
figure indicates that the expected scaling emerges for these quenches of free fermions, \ie  $\langle\bar{\psi}\psi\rangle_{ren} \sim \dt^{-(d-2)}$. Again, for $d=2$, the naive scaling formula no longer applies and rather a logarithmic scaling behaviour
is found. The holographic analysis of \cite{numer,fastQ} suggests a logarithmic enhancement should
arise in any even $d$. Unfortunately, fitting our numerical results is insensitive to these additional
logarithmic factors for $d\ge3$.  However, we can
demonstrate their presence analytically, as will be described in \cite{future}.

\noindent \textbf{4. Interacting Theories:} So far we showed that the 
early time scaling that first appeared in holographic
studies of quenches in strongly coupled theories \cite{numer,fastQ} also emerges for mass quenches in free field theories. This result then suggests that this scaling actually applies quite broadly and here we formulate a general argument for this
behaviour. Let us begin with a general quantum field theory of the form
\ben
S = S_\mt{CFT} + \int d^d x\, \lambda(t)\, \calo_\Delta(x)
\een
where $\calo_\Delta$ is a relevant operator with dimension $\Delta < d$. To describe the quenches similar to
those described above, we choose the coupling $\lambda(t)$ to have the form
\ben
\lambda(t) = \delta \lambda\, h(t/\delta t) \label{proc}
\een 
where $h(x)$ is a smooth function which rises from zero to $1$ and then returns to zero roughly over an
interval of width $\Delta x\sim1$ centered around $x=0$. 
Hence with this protocol \reef{proc}, the coupling $\lambda(t)$ experiences a finite pulse with an amplitude $\delta\lambda$ 
in an time interval of width $\dt$ centered around $t=0$. Since the theory is critical both in the past and future,
\ie outside of this interval, we can reliably 
calculate the expectation value of the operator using conformal perturbation theory, 
\bea
&&\!\!\!\!\langle \calo_\Delta(0)\rangle =
\langle \calo_\Delta(0) \rangle_\mt{CFT}- \delta \lambda \int d^d x\, h(t/\dt)\,  G_R (0,x) 
\nonumber\\
&&+ \frac{\delta\lambda^2}{2} \int d^d x\,h(t/\dt)\, d^dx'\,h(t'/\dt) \, K(x,x',0)
+\cdots
\label{gft}
 \eea
where all the expectation values here are evaluated in the critical theory $S_\mt{CFT}$.
In particular then, since we are considering
a relevant operator, the expectation value $\langle \calo_\Delta(0) \rangle_\mt{CFT}$ vanishes.
The second term above is the linear response
with the retarded correlator
\ben
G_R (0,x) = i \theta(t)\ \langle\,  [\calo_\Delta (x) , \calo_\Delta (0) ]\, \rangle_\mt{CFT}\,.
\een
Similarly, the third term involves a suitably ordered correlator of three insertions of $\calo_\Delta$ \cite{future}.
Of course, the expression (\ref{gft}) is UV divergent and we need to add counterterms to extract out the finite renormalized
expectation value. We will assume that this renormalization can be carried out to yield a result which only depends on
the two (renormalized) parameters, $\delta\lambda$ and $\dt$. Implicitly, we are assuming that the quench protocol does not lead
to any unconventional RG flows, as observed in \cite{eva,timedeprg}.

Now we observe that the final result is simply expressed in terms the renormalized parameters, 
$\delta \lambda$ and $\delta t$. However, since the
CFT correlators are independent of both of these parameters, 
$\dt$ must set
the scale for the integrals appearing in eq.~\reef{gft} and hence dimensional analysis demands
that the final finite result must take the form
\ben
\langle \calo_\Delta(0)\rangle_{ren} =
 a_1\, \delta\lambda\,\dt^{d-2\Delta} 
+ a_2 \, \delta\lambda^2\,\dt^{2d-3\Delta} + \cdots
\label{gft2}
\een
where the constants $a_n$ are finite numbers by assumption.  
Hence we see the linear response term produces the
desired scaling $\delta\lambda/ \dt^{2\Delta-d}$, however, there
is an infinite series of nonlinear contributions modifying this result.
At this point, we note that the above series is better expressed in terms of
a dimensionless effective coupling
\ben
g = \delta\lambda\, \dt^{d- \Delta}\,,\label{effect}
\een
in terms of which, eq.~\reef{gft2} becomes
\ben
\langle \calo_\Delta(0)\rangle_{ren}  =
 (\delta t)^{-\Delta} [\, a_1 g + a_2 g^2 + \cdots ]\,.
\label{stacker}
\een
Now for the fast quenches studied here, we hold
$\delta\lambda$ fixed while taking $\dt \rightarrow 0$ and therefore we are considering a limit
where the effective coupling \reef{effect} becomes vanishing small, \ie $g \rightarrow 0$ since $\Delta < d$.
Hence, in this limit, it is precisely the first term, with the desired scaling, which dominates the 
above expansion in powers of $g$ and the desired scaling will emerge for these pulsed quenches
irrespective of the details of the underlying CFT.

\noindent \textbf{5. Conclusions:} Recall that our counterterms were constructed using an adiabatic expansion. The physical intuition behind this
approach is that the high momentum modes do not care if the quench is fast or slow, as long as the quench rate is smaller than the scale of the cutoff. The quantum quenches considered in this Letter are fast compared to the scale of the relevant coupling, but slow compared to the scale of the cutoff. In fact, implicitly, we take the cutoff scale $\Lambda$ to infinity in renormalizing the physical observables. Therefore one would expect that the UV effects can be removed by subtracting the answer for slow quenches, which is of course the adiabatic expansion.
We have explicitly shown that this intuition is correct for free fields. The holographic calculation of \cite{numer,fastQ} indicates that this is valid for these strongly coupled theories as well. While we do not have a proof that this is also valid for arbitrary field theories, we believe that it is a reasonable assumption to make. It would be interesting to check if this is indeed true by explicit calculations in interacting field theories, \eg in large-$N$ vector models \cite{misha}. It would also be interesting to study
the early time scaling in quenches of systems where the cutoff $\Lambda$ remains finite. It is reasonable then to
expect that the scaling discussed here would only appear as an approximate description when $\Lambda\gg1/\dt\gg m$.

In the limit of `infinitely fast' quenches, our analysis revealed that various divergences appear for $\Delta>d/2$, 
suggesting that these processes can not actually be realized. While these results naively appear to contradict
the analysis of `instantaneous' quenches in \eg \cite{cc2,cc3,gritsev}, a more detailed examination revealed that
these are simply two different types of processes. In fact, the processes described in \cite{cc2,cc3,gritsev}
might better be thought of as the evolution of a certain far-from-equilibrium initial state with a fixed Hamiltonian.

To close, we showed in this Letter that both universal scaling of fast quenches and regularization of time dependent QFTs work even outside holography. It would be interesting to further generalize these results to arbitrary CFTs using linear response theory and try to understand the concrete relation between our quenches and other studies, 
\eg in the lattice or in weakly interacting theories. We hope to report on this in the near future.

\
\

\noindent {\bf Acknowledgements:} We would like to thank Joe Bhaseen, Robert Brandenberger, Alex Buchel, 
John Cardy, Anushya Chandran, Diptarka Das, Benjamin Doyon, Fabian Essler, Ganpathy Murthy, 
Anton van Niekerk and Krishnendu Sengupta for discussions. Research at Perimeter Institute is supported by the
Government of Canada through Industry Canada and by the Province of Ontario
through the Ministry of Research \& Innovation. RCM and DAG were also supported
by an NSERC Discovery grant. RCM is also supported by research funding
from the Canadian Institute for Advanced Research. 
The work of SRD is partially supported by the National Science Foundation grants NSF-PHY-1214341 and NSF-PHY-0970069. 
We thank the organizers and participants of the 2013 Austin Holography Workshop where this work was initiated.

\end{document}